\documentclass[twocolumn,aps]{revtex4}

\usepackage{graphicx}
\usepackage{dcolumn}
\usepackage{bm}
\usepackage[cmex10]{amsmath}

\begin{document}

\preprint{APS/123-QED}

\title{Influence of MgO tunnel barrier thickness on spin-transfer ferromagnetic resonance and torque in magnetic tunnel junctions}

\author{Witold Skowro\'{n}ski}
 \email{skowron@agh.edu.pl}
\author{Maciej Czapkiewicz}
\author{Marek Frankowski}
\author{Jerzy Wrona}
\author{Tomasz Stobiecki}
\affiliation{Department of Electronics, AGH University of Science and Technology, Al. Mickiewicza 30, 30-059 Krak\'{o}w, Poland
}

\author{G\"{u}nter Reiss}
\affiliation{Thin Films and Physics of Nanostructures, Bielefeld University, 33615 Bielefeld, Germany}

\author{Khattiya Chalapat}
\author{Gheorghe S. Paraoanu}
\affiliation{Low temperature laboratory, Aalto University, P.O.Box 15100, FI-02015 Aalto, Finland}
\author{Sebastiaan van Dijken}
\affiliation{NanoSpin, Department of Applied Physics, Aalto University School of Science, P.O.Box 15100, FI-00076 Aalto, Finland}

\date{\today}

\begin {abstract}
Spin-transfer ferromagnetic resonance (ST-FMR) in symmetric magnetic tunnel junctions (MTJs) with a varied thickness of the MgO tunnel barrier (0.75 nm $<$ $t_{\mathrm{MgO}}$ $<$ 1.05 nm) is studied using the spin-torque diode effect. The application of an RF current into nanosized MTJs generates a DC mixing voltage across the device when the frequency is in resonance with the resistance oscillations arising from the spin transfer torque. Magnetization precession in the free and reference layers of the MTJs is analyzed by comparing ST-FMR signals with macrospin and micromagnetic simulations. From ST-FMR spectra at different DC bias voltage, the in-plane and perpendicular torkances are derived. The experiments and free-electron model calculations show that the absolute torque values are independent of tunnel barrier thickness. The influence of coupling between the free and reference layer of the MTJs on the ST-FMR signals and the derived torkances are discussed.
\end{abstract}

\maketitle

\section{Introduction}
High density magnetic random access memories can be implemented using current-induced magnetization switching (CIMS) \cite{huai_observation_2004} which is caused by interactions between spin-polarized current and the magnetization of the free layer (FL) in magnetic tunnel junction (MTJ) cells. This phenomenon is called the spin-transfer-torque (STT) effect \cite{berger_emission_1996, slonczewski_current-driven_1996}. Moreover, STT is utilized in MTJ nano-oscillators that generate signals in the GHz frequency range \cite{petit_spin-torque_2007, deac_bias-driven_2008, skowronski_zero-field_2012}. In order to optimize MTJ parameters, so that they can compete with existing memory and microwave technologies, it is necessary to fully understand STT. The spin-torque diode effect enables quantitative measurements of STT parameters \cite{tulapurkar_spin-torque_2005, sankey_measurement_2007, kubota_quantitative_2007}.
In this work, we use the spin-torque diode effect to investigate the dependence of in-plane and perpendicular spin torques on MgO tunnel barrier thickness. 
The tunnel barrier determines the transport properties of the device, as it affects the tunneling magnetoresistance (TMR) ratio, the resistance area (RA) product and the coupling between the FL and the reference layer (RL). We show that the spin-torque ferromagnetic resonance (ST-FMR) spectra contain a double resonance mode for very thin MgO barriers due to strong ferromagnetic interlayer coupling. Moreover, the in-plane and perpendicular spin-torques do not depend on MgO barrier thickness, in agreement with free electron models \cite{wilczyski_free-electron_2008}.

\section{Experimental}
The MTJ stack with a MgO wedge tunnel barrier was deposited in a Singulus Timaris cluster tool system. The multilayer structure consisted of the following materials (thickness in nm): Ta(5) / CuN(50) / Ta(3) / CuN(50) / Ta(3) / PtMn (16) / Co$_{70}$Fe$_{30}$(2) / Ru(0.9) / Co$_{40}$Fe$_{40}$B$_{20}$(2.3) / wedge MgO(0.7 - 1.1) / Co$_{40}$Fe$_{40}$B$_{20}$(2.3) / Ta(10) / CuN(30) / Ru(7). The slope of the MgO wedge barrier was approximately 0.017 nm/cm. The deposition process was similar to the one used in our previous studies \cite{skowronski_interlayer_2010, wrona_low_2010}. After thin-film deposition, three different parts of the sample were selected for patterning into nanometer size pillars (later in the paper referred to as S1, S2 and S3, see Table ~\ref{tab:Static} for details). Using a three-steps electron beam lithography process, which included ion beam milling, lift-off and oxide and conducting layers deposition steps, nanopillars with an elliptical cross-section of 250 $\times$ 150 nm were fabricated. The pillars were etched to the PtMn layer. The electric leads to each MTJ nanopillar consisted of coplanar waveguides which were designed to match an impedance of 50 Ohms. To ensure good RF performance, the overlap between the top and bottom leads was about 4 $\mu$m$^2$, which resulted in a capacitance of less than 1 $\times$ 10$^{-14}$ F. Each set of MTJs with a constant MgO tunnel barrier consisted of 10 - 15 nanopillars.

\begin{table}[!t]
\renewcommand{\arraystretch}{1.3}
\caption{Summary of static parameters of the prepared MTJ nanopillars.}
\label{tab:Static}
\centering
\begin{tabular}{ccccc}
Sample No. & MgO thickness & TMR & RA product & Hs  \\
  & (nm) & (\%) & ($\Omega\mu$m$^2$) & (Oe)  \\
\hline
S1 & 1.01 & 170 & 9.6 & -21.7  \\
S2 & 0.95 & 165 & 6.24 & -3.7 \\
S3 & 0.76 & 110 & 2.86 & 47 \\
\end{tabular}
\end{table} 

\begin{figure}
\includegraphics[width=3.5in]{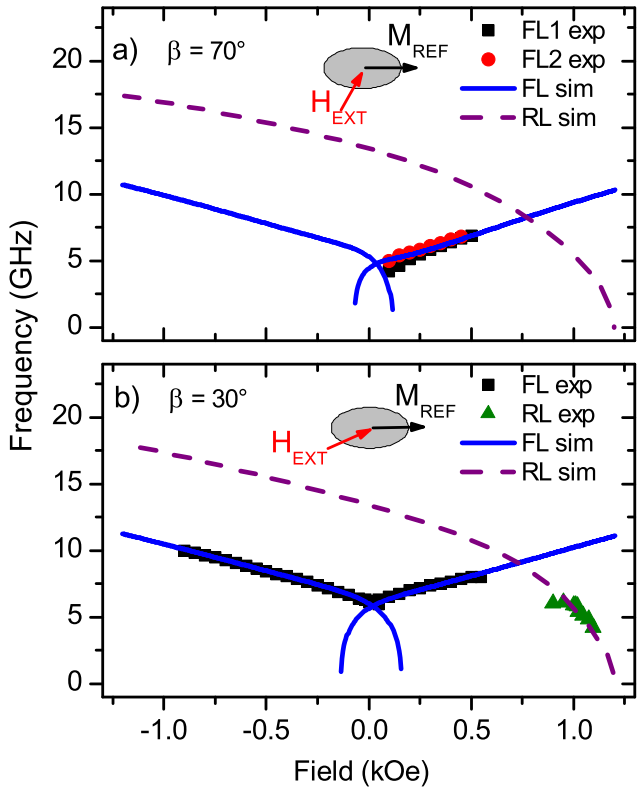}
\caption{The dispersion relation of sample S3 measured with the magnetic field applied at an angle of $\beta$ = 70$^\circ$ (a) and $\beta$ = 30$^\circ$ (b) with respect to the easy magnetization axis. The solid and dashed lines represent macrospin simulations of the FL and RL, respectively. (a) At an angle of $\beta$ = 70$^\circ$, the resonance frequency of two slightly separated FL modes increase with increasing magnetic field. (b) At an angle of $\beta$ = 30$^\circ$, magnetization precessions in both the FL and RL are measured.}
\label{fig:Kittel_1}
\end{figure}

ST-FMR measurement were conducted in a frequency range from 2 to 12 GHz. In these experiments, the application of an RF current to an MTJ generated a DC voltage (also called mixing voltage $V_{\mathrm{mix}}$) across the device, when the current frequency was brought into resonance with the resistance oscillations arising from the STT. The MTJs were placed in an in-plane magnetic field at an angle of $\beta$ = 70$^\circ$ with respect to the easy magnetization axis (except for the case presented in Fig. ~\ref{fig:Kittel_1}(b)), so that a large variety of angles $\theta$ between the junction's FL and RL could be obtained. We estimated $\theta$ from the assumption, that the resistance $R$ of the MTJ changes as follows:
\begin{equation}
\label{eq:TMR_cos}
\cos(\theta) =
\left(\frac{R_{AP}+R_P}{2}-R\right)\left(\frac{2}{R_{AP}-R_P}\right)
\end{equation} where \textit{R$_{AP}$} and \textit{R$_{P}$} are the resistance of the MTJ for an antiparallel and parallel alignment of the FL and RL magnetization, respectively.
In order to obtain the clearest STT results \cite{wang_bias_2009}, the strength and angle of the external magnetic field was adjusted so that magnetization of the FL is perpendicular to the magnetization of the RL ($\theta$ = 90$^\circ$). The magnitude of the RF input signal, connected to the MTJ through the capacitive lead of a bias tee, was fixed to -15 dBm. This resulted in a RF current (\textit{I$_{RF}$}) between 5 $\mu$A and 25 $\mu$A, depending on the sample resistance. \textit{I$_{RF}$} was calculated on the basis of the non-resonant background signal, using a model proposed in Ref. \cite{sankey_measurement_2007}. The bias voltage was fed through the inductive lead of the bias tee. $V_{\mathrm{mix}}$ was measured using a AC coupled lock-in amplifier, which was synchronized with the amplitude modulated signal from the RF generator. In this paper, positive bias voltage indicates electron transport from the bottom RL to the top FL.

\begin{figure}
\includegraphics[width=3.5in]{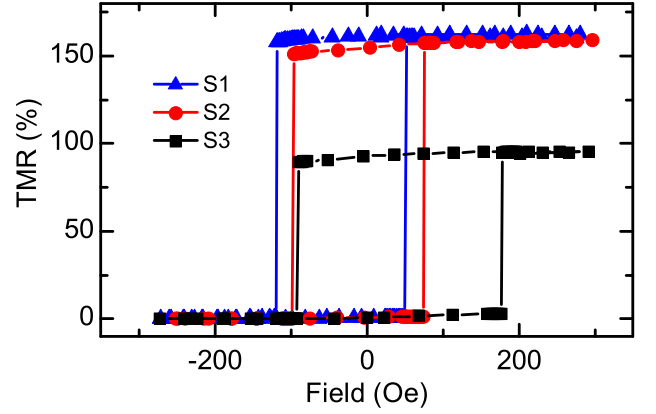}
\caption{TMR vs. magnetic field loops of samples S1-S3.}
\label{fig:TMR}
\end{figure}

\begin{figure}
\includegraphics[width=3.5in]{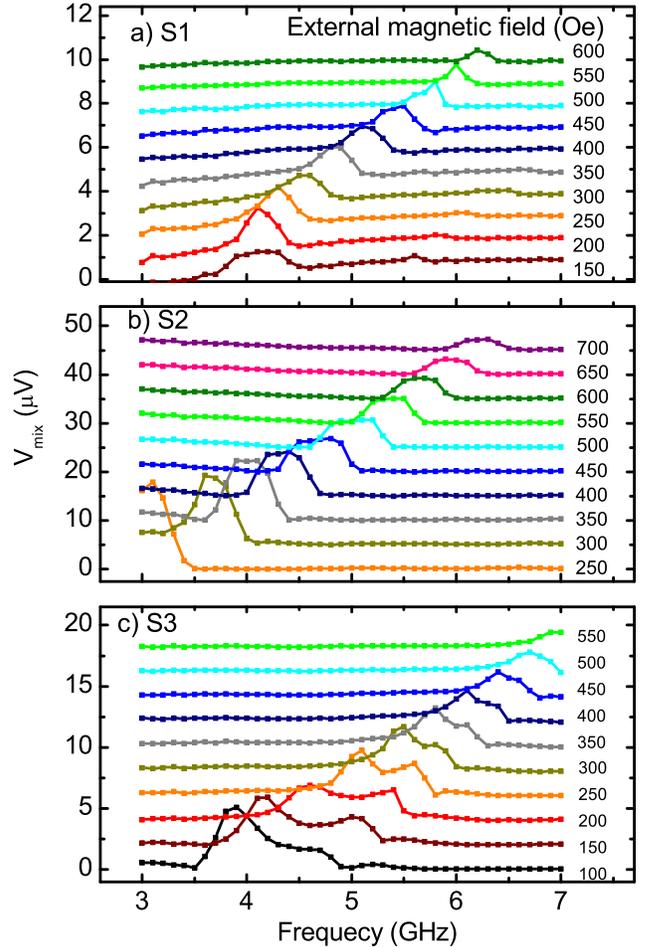}
\caption{ST-FMR spectra of samples S1 (a), S2 (b) and S3 (c) measured with various magnetic field applied at an angle of $\beta$ = 70$^\circ$ with respect to the easy magnetization axis. Only the RF signal (without DC bias voltage) was supplied to the MTJ. For sample S3 (c) two closely spaced peaks are visible.}
\label{fig:ST-FMR}
\end{figure}

\section{Results and discussion}
Table ~\ref{tab:Static} summarizes the TMR, the RA product and the static offset magnetic field (\textit{H$_S$}) for three sets of MTJs with different MgO tunnel barrier thickness. The representative TMR vs. magnetic field loops are presented in Fig. ~\ref{fig:TMR}. The high TMR ratio of 170\% for a 1.01 nm thick barrier and the exponential decrease in RA product with decreasing MgO thickness confirm good tunnel barrier quality \cite{skowronski_interlayer_2010}. Similar TMR ratios and RA products were measured on full wafers using a current in-plane tunnelling (CIPT) technique before patterning \cite{wrona_low_2010}. The overall offset field (\textit{H$_S$}) is shifted approximately 30 - 40 Oe with respect to the wafer-level measurements due to dipolar magnetostatic stray-field coupling in the nanopillar junctions. For the MTJ with a 1.01 nm thick tunnel barrier, antiferromagnetic stray-field coupling dominates the interaction between FL and RL (\textit{H$_S$} = -21.7 Oe). A reduction of the barrier thickness to 0.76 nm reverses the sign of the offset field (\textit{H$_S$} = 47 Oe). In this case, the FL and RL couple ferromagnetically due to direct interactions across the thin MgO tunnel barrier.

\subsection{ST-FMR}
Typical ST-FMR signals (without DC bias voltage) for samples S1 - S3 are presented in Fig. ~\ref{fig:ST-FMR}. We note that a single symmetric peak is measured for sample S2 in a wide magnetic field range. For this sample, the coupling between FL and RL is negligible. Moreover, the monotonic increase of the resonance frequency with applied magnetic field indicates that the FMR signal originates from magnetization precession in the FL \cite{cornelissen_free_2010}. A similar behavior is observed for sample S1, wherein the effective coupling between FL and RL is weakly antiferromagnetic. However, for sample S3, which is characterized by strong ferromagnetic coupling between FL and RL, an additional peak is measured. The origin of this double resonance mode is not entirely clear. In previous publications, it has been attributed to domain formation in the FL \cite{lee_excitations_2004}, higher-order spin wave excitations \cite{helmer_quantized_2010} and magnetization precession in other layers of spin-valve MTJs \cite{sankey_spin-transfer-driven_2006}. 
To analyze the double resonance mode in sample S3 in more detail, we performed macrospin simulations using the model presented in Ref. \cite{serrano-guisan_inductive_2011}. This model, based on the Stoner-Wolfarth approach, assumes coherent rotation of the FL and RL magnetization. By minimizing the system energy we find the angle of the FL and RL magnetizations with respect to the easy axis and on this basis, we calculate the dispersion relation. The simulated dispersion relations that are obtained for $\beta$ = 70$^\circ$ and  for $\beta$ = 30$^\circ$ are presented in Fig. \ref{fig:Kittel_1} together with the measured ST-FMR spectra. For $\beta$ = 30$^\circ$, the experimental and simulated FMR modes of the FL and RL are in good quantitative agreement. We note that the FMR signal of the RL is only measured when a large positive magnetic field is applied to the nanopillar junctions. The resonance frequency of the RL decreases with increasing field strength in this field range. The frequency of the double resonance peak in the spectra for $\beta$ = 70$^\circ$ (Fig. ~\ref{fig:Kittel_1}(a)), on the other hand, increase with applied field strength. The experimental dispersion relations now closely match simulated curve. Based on this analysis, we attribute the double resonance mode to inhomogeneous magnetization precession in the FL rather than FMR in the RL or any other magnetic layer of the MTJ stack. 

\begin{figure}
\includegraphics[width=3.5in]{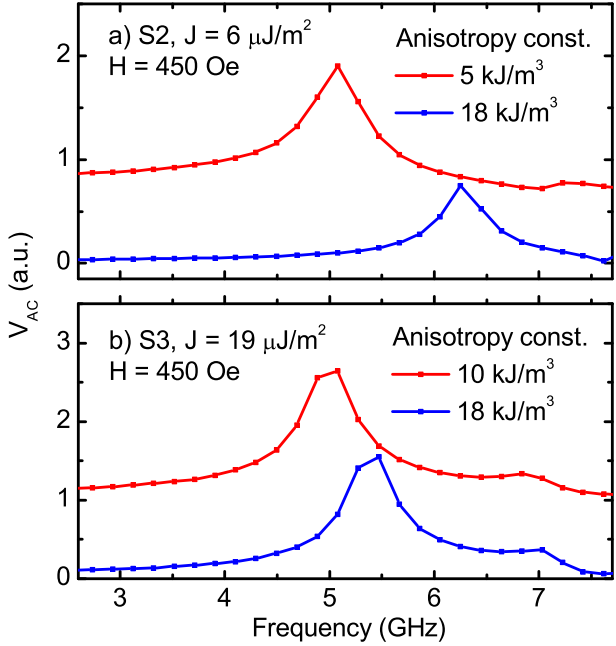}
\caption{Simulated ST-FMR curves for weak (a) and strong (b) ferromagnetic interlayer exchange coupling. In \textsc{oommf} simulations, a voltage step was used to excite magnetization precession in the FL of a MTJ structure. The dimensions of the simulated and experimental junctions are identical. The existence of closely-spaced double-peak ST-FMR signal for strong coupling is independent of the anisotropy constant. }
\label{fig:st-fmr-sim}
\end{figure}

To further elucidate the origin of double-mode FL spectra, we simulated the resonance characteristics of MTJ nanopillars using \textsc{oommf} software \cite{donahue_oommf_1999} with an additional extension enabling calculations of TMR and STT effects \cite{frankowski_www}. In these micromagnetic simulations, elliptical multilayer systems with a 2 nm thick FL, a 1 nm thick MgO tunnel barrier, a 2 nm thick high-anisotropy RL, antiferromagnetically coupled to a 2 nm thick exchange-biased pinned layer (PL), were used. 
The area of the junction was identical to the experimental structures. The interlayer exchange coupling and anisotropy energies were experimentally determined by magnetic and magnetotransport measurements. Variation of the ferromagnetic interlayer exchange coupling from 0 to 19 $\mu$J/m$^2$ in the simulations yielded results comparable to the experimental data. We note that dipolar coupling between the FL and RL is intrinsically calculated and taken into account in \textsc{oommf}. Thus depending on the strength of the interlayer exchange coupling (input parameter), the effective coupling between FL and RL varies from antiferromagnetic to ferromagnetic in accordance with the experimental results on samples S1 - S3. 

The dynamic simulations were conducted in the following way: first, an external magnetic field was applied at an angle with respect to the magnetic easy axis. After relaxation, a voltage step was applied to exert a STT on the FL. The voltage step amplitude was adjusted, so that the FL magnetization oscillations changed the MTJ resistance by a few Ohms. The used values correspond to an AC current of a few $\mu$A, which closely mimic the experimental conditions and ensures that the magnetization oscillations are within the linear regime. Finally, the resonance spectra were obtained by Fourier transformation of the time-derivative damped oscillation of the simulated tunneling magnetoresistance.

Figure \ref{fig:st-fmr-sim} presents the simulated ST-FMR spectra for two MTJ nanopillars that closely resemble experimental samples S2 and S3. The simulations confirm that the magnetization of the RL does not precess under these conditions ($\beta$ = 70$^\circ$) in the investigated frequency range. 
For a weak interlayer exchange coupling energy of $J$ = 6 $\mu$J/m$^2$ (sample S2), a single resonance peak is simulated for different FL anisotropy energies - Fig. \ref{fig:st-fmr-sim}a - and different magnetic field strength (not shown), which fulfills the Kittel dispersion relation. 
For a larger ferromagnetic coupling energy of $J$ = 19 $\mu$J/m$^2$ (sample S3), an additional broad resonance peak was resolved in the simulations (Fig. ~\ref{fig:st-fmr-sim}(b)), regardless of the FL magnetic anisotropy. This behavior is reminiscent to the experimental behavior of sample S3 with a 0.76 nm thin MgO tunnel barrier. The simulations thus confirm that the the double resonance mode originates from inhomogeneous magnetization precession in the FL of the MTJ nanopillar stack due to strong interlayer exchange coupling between FL and RL. 

\subsection{Torques and torkances}
In order to obtain the STT components, i.e., in-plane torque $\tau_{\parallel}$ and perpendicular torque $\tau_{\perp}$, from the ST-FMR measurements, we used the model presented in Ref. \cite{wang_bias_2009}. Here, we assume a simplified formula for $V_{\mathrm{mix}}$:
\begin{subequations}
\label{eq:Vmix}
\begin{align}
V_{\mathrm{mix}} &= \frac{1}{4}\frac{\partial^2V}{\partial I^2}I_{\mathrm{RF}}^2 \\
& +\frac{1}{2}\frac{\partial^2V}{\partial I \partial\theta}\frac{\hbar \gamma \sin\theta}{4eM_S Vol \sigma}I_{\mathrm{RF}}^2 [\xi_{\parallel}S(\omega) - \xi_\perp \Omega A(\omega)],
\end{align}
\end{subequations} where $\hbar$ is the reduced Planck's constant, $\gamma$ is the gyromagnetic ratio, \textit{e} is the electron charge, \textit{Vol} is the volume of the FL, \textit{M$_S$} is the saturation magnetization of the FL, $\sigma$ is the linewidth, $\xi_{\parallel}$ = 2(\textit{e/$\hbar$}sin$\theta$)(\textit{d}V/\textit{d}I)\textit{d}$\tau_{\parallel}$/\textit{d}V and  $\xi_{\perp}$ = 2(\textit{e/$\hbar$}sin$\theta$)(\textit{d}V/\textit{d}I)\textit{d}$\tau_{\perp}$/\textit{d}V are the magnitudes of the symmetric \textit{S}($\omega$)=[1+($\omega$-$\omega_m)^2$/$\sigma^2$]$^{-1}$ and asymmetric \textit{A}($\omega$)=[($\omega$-$\omega_m)$/$\sigma$]\textit{S}($\omega$)  lorentzians components, and $\Omega_\perp$=$\gamma$\textit{N$_x$}\textit{M}$_{\mathrm{eff}}$/$\omega_m$, \textit{N$_x$}=4$\pi$+(\textit{H$_z$}-\textit{H$_a$}sin$^2\beta$)/\textit{M}$_{\mathrm{eff}}$, where $\omega_m$ is the resonant frequency, \textit{H$_z$} is the sum of the applied external magnetic field and the offset field acting on the precessing FL, \textit{H$_a$} is the in-plane anisotropy field of the FL and 4$\pi M_{\mathrm{eff}}$ is the effective out-of-plane anisotropy of the FL. We neglected the terms (2c) - (2g) of Ref. \cite{wang_bias_2009} because in our case $\theta$ $\approx$ 90$^\circ$.

\begin{figure}
\centering
\includegraphics[width=3.5in]{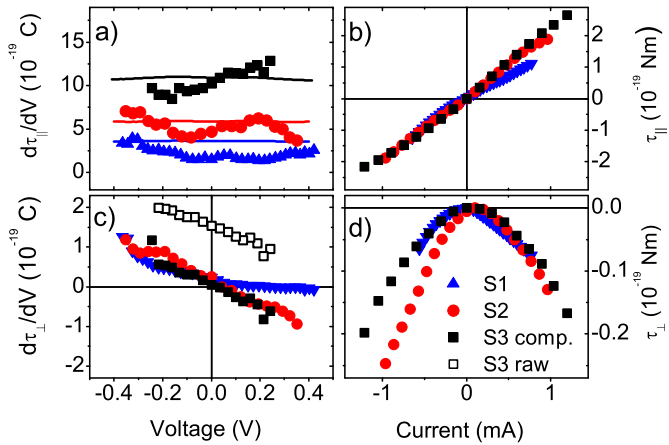}
\caption{Bias dependence of the in-plane torkance (a), in-plane torque (b), perpendicular torkance (c) and perpendicular torque (d) for MTJs with different MgO barrier thickness. The solid lines in (a) represent calculations based on Eq. ~\ref{eq:Torkance_thoery}. The torque values are numerically integrated from experimentally determined torkances. $\tau_{\perp}$ for sample S3 was compensated for an error originating from asymmetric ST-FMR resonances.}
\label{fig:Torques_sum_1}
\end{figure}

Figure \ref{fig:Torques_sum_1}a presents a comparison of the in-plane torkance in samples S1, S2, and S3. The absolute value of the in-plane torkance increases with decreasing barrier thickness and it only weakly depends on DC bias voltage. According to Slonczewski's free electron model for elastic tunneling in symmetric MTJs, the in-plane torkance is proportional to the differential conductance measured for parallel alignment of FL and RL  \cite{slonczewski_theory_2007}:

\begin{equation}
\label{eq:Torkance_thoery}
\frac{d\tau_{\parallel}}{dV} =
\frac{\hbar}{2e}\frac{2p}{1+p^2}\left(\frac{dI}{dV}\right)_\parallel
\end{equation} 

By using Julliere’s model to derive the spin polarization of the tunneling current $p$ at V = 0 V, we found a good match between our experimental data and theoretical calculations based on Eq.\ref{eq:Torkance_thoery} (Fig. \ref{fig:Torques_sum_1}(a)). The absolute torque values in Fig. \ref{fig:Torques_sum_1}(b) were obtained by numerical integration of the data in Fig. \ref{fig:Torques_sum_1}(a). Obviously, the in-plane torque varies linearly with DC bias current and it is independent of MgO tunnel barrier thickness. These results are in good agreement with previously published experimental data in Refs \cite{sankey_measurement_2007, kubota_quantitative_2007, wang_bias_2009, wang_time-resolved_2011} and calculations based on an \textit{ab initio} approach \cite{heiliger_ab_2008, jia_nonlinear_2011}.

Experimental data on the perpendicular torkance are summarized in Fig. \ref{fig:Torques_sum_1}(c). For samples S1 and S2, the torkance decreases with DC bias voltage and $d\tau_{\perp} / dV $ = 0 for zero DC bias voltage as predicted by theoretical calculations. However, a discrepancy is observed for sample S3. In this sample, strong ferromagnetic coupling between the FL and RL of the MTJs results in asymmetrical double resonance modes in the ST-FMR spectra. The fitting procedure based on Eq. \ref{eq:Vmix} therefore introduces an error in the experimental torkance values for this sample. A good match with theoretical calculations is obtained when this artifact is compensated by subtraction of a constant torkance value. Figure \ref{fig:Torques_sum_1}(d) illustrates that the absolute perpendicular torque varies quadratically with DC bias current. Moreover, $\tau_{\perp}$ is similar for all samples. We note that different torque versus bias dependencies have been measured recently. Especially, it has been shown that the shape of $\tau_{\perp}$(V) curves can change from quadratic to linear \cite{chanthbouala_vertical-current-induced_2011, matsumoto_spin-torque_2011}. However, such effects were only measured in asymmetric MTJs with different FL and RL electrodes. In our junctions, the composition and thickness of the CoFeB electrodes are the same

\section{Summary} 

In summary, we have investigated MTJ nanopillars with varied MgO tunnel barrier thickness using the spin-torque diode effect. We measured a symmetric ST-FMR signal for samples with $t_{\mathrm{MgO}}$ $>$ 0.9 nm. In this case, the coupling between FL and RL is weakly antiferromagnetic. Contrary, double and closely-spaced resonance modes were obtained for MTJs with a 0.76 nm thick tunnel barrier. Macrospin and micromagnetic simulations indicate that the asymmetric double-peaks originate from inhomogeneous magnetization precession in the FL caused by ferromagnetic coupling to the RL. The in-plane and perpendicular torques scale with DC bias current and they are independent of MgO tunnel barrier thickness.”

\section*{Acknowledgement}
We would like to thank Singulus Technologies AG for consultation and technical help with MgO wedge MTJs preparation. We also acknowledge help of Micha\l{} Wilczy\'{n}ski, Piotr Ogrodnik and Renata \'{S}wirkowicz for a fruitful discussion regarding STT models. Project supported by the Polish National Science Center grant 515544538, Polish Ministry of Science and Higher Education Diamond Grant DI2011001541 and Swiss Contribution by NANOSPIN PSPB-045/2010  grant. W.S and T.S. acknowledges the Foundation for Polish Science MPD Programme co-financed by the EU European Regional Development Fund. G.R. acknowledges support from the DFG (contract RE 1052/21-1). G.S.P and K.C. acknowledge support from the Commission of Higher Education of Thailand and Academy of Finland (nos. 129896, 118122, and 135135). S.v.D. acknowledges financial support from the Academy of Finland (grant no. 127731).

\bibliographystyle{nature}

\end{document}